# Comparing Missing Data Methods for Estimating Average Treatment Effects Under Time-Varying Confounding: A Simulation Study

*Running title: Missing Data Methods in Causal Inference*

Ben Swallow[1], Lars Brestrich[1], Victor Velasco-Pardo[2]

[1]*School of Mathematics and Statistics, University of St Andrews, St Andrews, United Kingdom*

[2]*School of Genetics and Cancer, University of Edinburgh, UK*

Correspondence: Ben Swallow, School of Mathematics and Statistics, University of St Andrews, St Andrews, KY16 9SS, United Kingdom. Email: bts3@st-andrews.ac.uk

**Keywords:** average treatment effect; causal inference; missing data; multiple imputation; propensity score weighting; simulation study

**Funding information:** None

## Abstract

Missing data and confounding are common in real-world statistical applications, yet few studies have examined how imputation methods perform under time-varying confounding in binary variables, or how missingness mechanism, missing rate, missingness location and sample size jointly affect performance and the underlying identifiability conditions. We generated synthetic data and conducted a simulation study comparing missing data methods across scenarios varying these factors. Missingness was introduced in both treatment and outcome variables, and we applied stratified hot deck imputation, single mode imputation, multiple imputation with chained equations (MICE), and complete-case analysis. Average treatment effect (ATE) estimates were obtained using logistic regression with propensity score weighting, and we measured coverage, absolute bias and empirical standard errors across 48 scenarios with 500 replications each. Performance was primarily driven by the missingness mechanism and choice of method, with multiple imputation generally achieving better coverage and lower bias than other methods. Missingness location was also important, while missing rate and sample size primarily affected positivity violations, which were most pronounced under MNAR, high missingness and low sample sizes. Exchangeability violations from mild to moderate confounding were adequately controlled for by propensity score models, whereas strong confounding produced a modest decrease in coverage. Further research should examine additional ways identifiability conditions can be violated under missingness, using more advanced methods and more complex missingness scenarios.

## 1 | Introduction

Missing data is common in real-world statistical applications, and in causal inference settings subjects frequently lack values for treatment, outcome, or confounding variables, most often through loss to follow-up, dropout, or non-response.[1] Because standard statistical methods generally assume complete data, restricting analysis to complete cases is the simplest way to circumvent missingness, but this approach commonly produces biased and inefficient estimates.[2,3] More elaborate missing data methods instead attempt to reconstruct a complete pseudo-population so that treatment effects can be estimated as though all subjects were fully observed.[1]

Unbiased causal inference relies on identifiability conditions - exchangeability, positivity and consistency - that can be unrealistic in cases of insufficient data.[1] Exchangeability requires that potential outcomes be independent of treatment assignment (or independent conditional on measured covariates in observational data); positivity requires a non-zero probability of receiving each treatment level within every stratum of the covariates; and consistency requires that the observed outcome under a given treatment equal the corresponding potential outcome. These conditions can be threatened directly by missing data and are further complicated by confounding, since accurately capturing the relationship between confounders, treatment and outcome is essential for adjustment, and missing data can make this more difficult.

A common approach for confounding adjustment is propensity score weighting, in which subjects are weighted by the inverse of their estimated probability of receiving the treatment they in fact received, given measured confounders.[4] This approach can restore exchangeability provided all confounders are measured, but if missing data methods fail to preserve the underlying relationships between treatment, outcome and confounders, propensity score models may no longer adjust effectively.

Many methods exist for both handling missing data and adjusting for confounding, and there is a well-known methodological overlap between the two problems: missing data can itself be understood as a form of selection bias, and inverse-probability weighting is common to both fields.[1] However, studies that jointly examine missingness and confounding parameters are scarce. Simulation studies comparing imputation methods for propensity score analysis of binary outcomes under MCAR and MAR have found that multiple imputation achieves the best statistical properties, although mean imputation can also perform reasonably well.[5] Comparisons of listwise deletion and multiple imputation under varying missingness rates have shown that multiple imputation consistently outperforms listwise deletion, particularly as missingness increases, and that the number and location of missing covariates has comparatively little effect on bias.[6] Simulation-based bias analyses of unmeasured confounding have shown that greater confounding strength increases bias, but that measured confounding can be effectively addressed with standard adjustment methods such as propensity scores.[7] A study focusing specifically on missingness in confounders, evaluated under three simulation settings, found that complete-case analysis and multiple imputation perform well only in the absence of unmeasured confounding, that missing-indicator methods are not recommended in isolation, and that no single method performs well across all scenarios.[8] None of these studies, however, jointly varies missingness mechanism, missing rate, missingness location, confounding strength and sample size, or explicitly links performance to violations of the underlying identifiability conditions. There is also little in the literature on accounting for missing data when the mechanism is assumed not at random[9].

We conducted a simulation study to compare the performance of complete-case analysis, single mode imputation, stratified hot deck imputation and multiple imputation by chained equations (MICE)[10] for estimating the average treatment effect (ATE) under a full-factorial combination of missingness mechanism, missing rate, missingness location (treatment vs. outcome), confounding strength and sample size, benchmarked against a full-data model. Average treatment effects were estimated using logistic regression with inverse-probability treatment weighting (IPTW), and performance was assessed via absolute bias, 95% confidence interval coverage and empirical standard error, alongside the prevalence of extreme propensity scores as an indicator of positivity violations.

## 2 | Methods

### *2.1 | Simulation design*

We used a full-factorial simulation design in which each unique combination of parameters (described below) constituted a scenario. Each scenario was replicated 500 times to obtain stable performance estimates, following recommended practice for simulation studies.[11] Each replication (i) generated a synthetic longitudinal dataset, (ii) fitted a full-data benchmark model, (iii) introduced missingness according to the scenario, (iv) applied complete-case analysis, single mode imputation, hot deck imputation and MICE, (v) fitted a logistic regression model with IPTW after each method, and (vi) recorded model output, including the occurrence of extreme propensity scores. Performance measures were then calculated for each scenario (Section 2.5).

### *2.2 | Data generation model*

Synthetic datasets were generated using the longitudinal data-generating process of Velasco-Pardo et al, which simulates time-varying treatment and survival processes under time-varying confounding in a vaccine-efficacy setting, with variable distributions informed by population data from a large UK primary-care multimorbidity study.[12,13] Each dataset contained three treatment time points, generating binary treatment indicators $A_0$, $A_1$, $A_2$ and binary survival outcomes $Y_0$, $Y_1$, $Y_2$, $Y_3$, together with four confounders: number of comorbidities, age group, area-level deprivation (SIMD), and sex.

$$A_k \sim \text{Bin}\left(1, \text{expit}\left(\gamma_{k0} + \sum_{q=1}^{18} \gamma_{kq} Z_q + \sum_{0 \le l < k} \gamma_{k,19+l} A_l\right)\right), \quad k = 0, 1, 2$$

Each treatment $A_k$ affects the subsequent outcome $Y_{k+1}$ and subsequent treatment $A_{k+1}$; each outcome $Y_k$ affects subsequent treatment; and confounders affect all treatment and outcome probabilities (Figure 1). We focused on missingness at the final observed time point (k=2), the setting most likely to introduce clinically relevant selection bias. Treatment allocations following a structural regimen (e.g., subjects who did not survive cannot receive further treatment) were treated as pre-existing structural missingness, distinct from the experimentally introduced missingness mechanisms; missingness was applied to all covariates regardless of structural-missingness status, and structural missingness was reinstated afterwards. Table 1 summarises the marginal distributions of treatment, outcome and confounder variables for a representative scenario (n=1000, ρ=-0.4); outcome variables were markedly more sparse than treatment variables, a feature relevant to the interpretation of our results (Section 4.2).

### *2.3 | Missingness Generation*

Missingness was introduced after data generation following the approach of Section 2.2. The missing data mechanisms (MCAR, MAR and MNAR) and the rate of missingness were varied independently across scenarios in order to measure their effect on the validity of downstream analysis. Each missingness mechanism is associated with two sets of variables. The first is a set of target variables where missingness will be introduced. Under MAR and MNAR, there is an additional set of predictor variables, i.e. the variables that the missingness in the targets depends on. Under MNAR, missingness depends on the variables themselves, so target and predictors were identical.

Missingness in each variable was set separately and exhibits the same missing rate defined in the respective scenario. Pseudocode for each mechanism is provided in the Supporting Information.

#### Generating MCAR

Generating MCAR mechanisms should be generated such that each subject has the same probability of being missing[14]. Our implementation follows one of the described methods, which simply selects random values of the variable which are then set to a null value.

#### Generating MAR

There are four distinct types of methods to introduce MAR, including single and multiple cutoff, percentile methods and regression methods[14]. Due to the categorical nature of the data, percentile methods are not appropriate and would create very biased results, especially for binary variables. Regression models are computationally expensive, and one cannot easily set the rate of missingness. Therefore, a single cutoff method adapted for categorical data was used, where the missingness for each respective level of predictor variable was introduced.

The single cutoff method is recommended as a valid method[14], especially when trying to create strong missingness relationships and maximise the difference between MAR and MCAR.

#### Generating MNAR

Under MNAR, the probability of missingness in a variable depends on an unobserved value, which can include the variable's own underlying missing value. Therefore, we can simply re-use the same logic as for MAR generation and set our predictor variable to be the same as the target variable[14], which will result in valid MNAR.

## *2.4 | Parameters*

We varied five parameters. Confounding strength (ρ) was set to -0.1, -0.4 or -0.7, reflecting a plausible range for medical research. Sample size (n) was 1000 or 5000; preliminary screening showed unstable performance below n=1000 and negligible differences between n=5000 and n=10,000. Missing rate was 20% or 50%, since preliminary screening indicated that rates below 20% produced negligible between-scenario differences. Missingness mechanism was MCAR, MAR with missingness in treatment variables (MAR(A)), MAR with missingness in outcome variables (MAR(Y)), or MNAR; missingness was introduced in treatment variables $A_1$ and $A_2$ or outcome variables $Y_1$ and $Y_2$ using a single-cutoff approach adapted for categorical data, following recommended approaches for generating missingness in simulation studies.[14] Under MNAR, the target and predictor variables for missingness were identical.

## *2.5 | Missing data methods*

Complete-case analysis restricted the analytic sample to subjects with fully observed data.[2] Although simple, this approach is unbiased only under MCAR and can otherwise introduce selection bias while discarding information and reducing effective sample size.[15] Full pseudocode for the methods are provided in the Supplementary Information.

Single mode imputation replaced missing values with the most frequently observed (modal) value of the variable - the categorical analogue of mean imputation.[2]

Stratified hot deck imputation grouped subjects into strata defined by up to six fully observed covariates and imputed missing values using a randomly selected complete "donor" from the same stratum, with random-value fallback where no matching donor or grouping variable was available.[16] Hot deck methods are considered unbiased under MCAR and MAR provided covariate information is available for both responding and non-responding units, though single imputation approaches generally understate uncertainty, since they yield only one imputed value per missing entry.[2.17]

MICE[9] generated m=5 imputed datasets by iteratively regressing each incomplete variable on all other variables using logistic regression and drawing imputed values stochastically until convergence. Each imputed dataset was analysed separately, and estimates were combined using Rubin's rules:[18] the pooled estimate is the average of the m estimates; the total variance combines the within-imputation variance (the average of the m variance estimates) and the between-imputation variance (the variance of the m estimates across imputations, inflated by a factor of 1+1/m).[15.19] Multiple imputation is approximately unbiased under MAR, provided the imputation model is correctly specified.[15]

### *2.6 | Simulation analysis*

#### Pooling Estimates

The m estimates obtained are combined into an overall estimate and variance-covariance matrix using Rubin's rules.[15.18]

Estimates are pooled using:

$$\bar{Q} = m^{-1} \sum_{\ell} \hat{Q}^{(\ell)}$$

Between-imputation variance is calculated using:

$$B = (m-1)^{-1} \sum_{\ell} \left( \hat{Q}^{(\ell)} - \bar{Q} \right)^2$$

Within-imputation variance is calculated using:

$$\bar{U} = m^{-1} \sum_{\ell} U^{(\ell)}$$

Finally, the estimated total variance is:

$$T = \left(1 + m^{-1}\right) B + \bar{U}$$

### *2.7 | Modelling and performance measures*

For each scenario and missing data method, we estimated the ATE using a logistic regression model of the outcome on treatment, weighted by inverse-probability treatment weights derived from a propensity score model, with robust (HC3 sandwich) standard errors, following the modelling approach of Velasco-Pardo et al.[12] We assessed performance using absolute bias, 95% confidence interval coverage and empirical standard error, calculated together with their Monte-Carlo standard errors,[20] following recommended practice for simulation studies.[11] We additionally recorded the proportion of extreme propensity scores ($<0.05$ or $>0.95$) as an indicator of positivity violations, and excluded scenario replications with extreme ATE estimates ($|\hat{A}|>100$) from the parameter analysis as numerically unstable.

### *2.8 | Statistical analysis*

We compared marginal method performance graphically and used repeated-measures ANOVA within a linear-model framework to quantify the contribution of each parameter and its first-order interactions to variance in bias, coverage and empirical SE, using partial eta-squared ($\eta^2$) as the effect-size measure. Because ANOVA based on simulation summary statistics is typically over-powered,[6] we used $\eta^2>0.134$ (conventionally a "large" effect) with $p<0.001$ as our threshold for practical significance, while also discussing medium effects ($\eta^2>0.06$). Where overall differences were significant, we used Tukey's honestly significant difference (HSD) test for pairwise comparisons between parameter levels.

## 3 | Results

### *3.1 | Extreme propensity scores and exclusions*

Extreme propensity scores were concentrated in MNAR scenarios and at 50% missingness (Table 2). Across mechanisms, MNAR scenarios showed a mean of 23.6% of weights with extreme propensity scores, compared with well under 5% for MCAR, MAR(A) and MAR(Y). Overall, 20% missingness yielded a mean extreme-score proportion of 5.1%, versus 10.1% at 50% missingness; smaller samples (n=1000) also showed a modestly higher proportion than larger samples (n=5000), while confounding strength and missingness location had little effect. Single mode imputation showed a consistently higher rate of extreme propensity scores than the other three methods.

Exclusion rates due to unstable ATE estimates followed the same overall pattern but were generally lower than the corresponding rates of extreme propensity scores, and were more strongly affected by sample size; multiple imputation exclusion rates were negligible throughout, likely reflecting its use of several fitted models per replication.

### *3.2 | Marginal method performance*

For the $A_1$ ATE estimate (Figure 2), multiple imputation achieved the lowest bias and highest coverage of all methods, closely approaching the full-data benchmark, with empirical SE comparable to hot deck imputation. Hot deck achieved approximately 90% marginal coverage but with substantial variability, and outperformed both single mode imputation and complete-case analysis on empirical SE. Single mode imputation and complete-case analysis performed similarly to one another, with complete-case analysis showing somewhat lower bias and higher coverage, but larger empirical SE. Results for the $A_2$ estimate were qualitatively similar.

### *3.3 | Sources of variance*

ANOVA models explained more than 70% of variance in each performance measure (Table 3). For bias, only missingness mechanism reached the pre-specified threshold for practical significance ($\eta^2=0.140$); method, sample size and their interactions with mechanism contributed medium-sized effects. For coverage, method ($\eta^2=0.175$) was the only parameter to reach practical significance, with confounding strength showing a borderline large effect ($\eta^2=0.129$); mechanism, sample size, and their interactions with confounding strength contributed further. For empirical

SE, both method ($\eta^2$=0.192) and mechanism ($\eta^2$=0.181) reached practical significance, with sample size and the mechanism-by-method interaction showing borderline large effects.

### *3.4 | Method performance by parameter*

Smaller samples (n=1000) were associated with higher bias and empirical SE than larger samples (n=5000, Figure 3), particularly for single mode imputation and complete-case analysis; multiple imputation and the full-data model were comparatively unaffected. Mean coverage was, on average, 11 percentage points higher at n=1000 than n=5000 (Table 4), although this appeared to reflect inflated robust standard errors under small-sample instability rather than genuinely better inference, since model-based standard errors showed the opposite pattern.

Increasing confounding strength had little effect on bias but was associated with reduced coverage, especially between $\rho$=-0.4 and $\rho$=-0.7 (Figure 4, Table 4); this reduction was most pronounced for complete-case analysis and single mode imputation, while multiple imputation was least affected.

Bias was lowest under MCAR and increased progressively under MAR(A), MAR(Y) and MNAR, with MNAR producing significantly higher bias and empirical SE than all other mechanisms for every method (Figure 5, Table 4). Differences in coverage between mechanisms did not reach statistical significance overall, though multiple imputation maintained the most stable coverage across mechanisms, including MNAR, while single mode imputation performed particularly poorly under MAR(Y).

A missing rate of 50% was associated with significantly higher bias and empirical SE than 20% (Figure 6, Table 4), with multiple imputation comparatively unaffected; the effect of missing rate on coverage was not statistically significant.

### *3.5 | Best- and worst-performing scenarios*

Across methods, the best performance (lowest bias, highest coverage, lowest empirical SE) was generally observed under MCAR or MAR(A), low missing rates and larger samples, while the worst performance occurred under MNAR, high missing rates and small samples (Table 5). Complete-case analysis and single mode imputation additionally showed markedly reduced coverage under strong confounding even within MCAR scenarios.

## 4 | Discussion

### *4.1 | Method performance*

Method choice was a strong determinant of coverage and empirical SE, and contributed materially to bias, underscoring that the choice of missing data method is central to reliable ATE estimation under missingness. MICE consistently outperformed the alternative methods, in agreement with previous simulation studies of multiple imputation;[5,8,6] it produced low bias, high coverage and low empirical SE in most scenarios, and unlike the other methods examined here, remained comparatively robust under MNAR, consistent with reports that multiple imputation can perform surprisingly well outside its formal assumptions.[21]

Complete-case analysis showed low bias but poor and highly variable coverage outside MCAR, consistent with the expectation that non-random missingness introduces selection bias under complete-case restriction.[15] This is broadly consistent with reports that complete-case analysis can perform comparably to multiple imputation under low missingness and large samples;[5] in our results, complete-case analysis was more strongly degraded by confounding strength, which likely explains the more pessimistic picture here.

Single mode imputation performed poorly throughout, particularly under MAR, likely because modal imputation of binary variables introduces substantial distortion of the variable distribution and its relationship to other covariates, consistent with previous findings that mode imputation of dichotomous variables underperforms other methods, including multiple imputation.[22] Its probable underestimation of standard errors[2] likely compounds its poor coverage.

Stratified hot deck imputation performed better than complete-case analysis and single mode imputation, especially under MCAR and MAR(A), consistent with theoretical expectations that hot deck imputation is unbiased under these mechanisms,[16] but its coverage remained inferior to multiple imputation throughout, and it should not be preferred over multiple imputation where feasible.

### *4.2 | Parameter effects*

Extreme propensity scores, an indicator of positivity violations, were concentrated under MNAR and high missing rates, and to a lesser extent under small samples, consistent with sparse covariate-treatment strata under these conditions.

Confounding strength primarily affected coverage rather than bias, with strong confounding ($|\rho|=0.7$) substantially reducing coverage, especially for complete-case analysis and single mode imputation. This suggests that propensity score models were less able to restore conditional exchangeability under strong confounding when missing data methods failed to preserve the underlying variable relationships needed for adjustment.

Missingness location (treatment vs. outcome) mattered: coverage for multiple imputation and hot deck decreased when missingness affected treatment rather than outcome variables, likely reflecting the highly skewed distribution of the (largely positive) outcome variables, which limits the information available for imputation of rare events; this contrasts with reports that missingness location itself has little impact,[22] and suggests that the effect of missingness location may depend on the underlying variable distribution.

Missingness mechanism, and its interaction with method, was the strongest driver of bias and empirical SE, consistent with the expectation that MCAR produces the least bias, MAR is unbiased if appropriately modelled, and MNAR produces biased estimates unless explicitly modelled.[2] Notably, multiple imputation maintained relatively high coverage even under MNAR by inflating its standard errors to reflect the additional uncertainty, rather than by reducing bias.

Smaller samples were associated with higher bias and empirical SE, most likely reflecting finite-sample instability in variance estimation[23] rather than a true small-sample bias in the estimators; this was corroborated by higher exclusion rates for unstable estimates at n=1000. Coverage estimates at small sample sizes should therefore be interpreted cautiously, since the direction of the sample-size effect on coverage differed between robust and model-based standard errors.

Higher missing rates (50% vs. 20%) increased bias and empirical SE for all methods except multiple imputation, consistent with reports of substantial bias increases at high missingness,[6] in contrast with other work reporting only modest degradation;[8] this discrepancy may reflect differences in the underlying data-generating processes and imputation methods used across studies.

### *4.3 | Limitations*

This study is subject to several limitations. Missingness was introduced via a single mechanism at a time and in a single set of variables, which is unlikely to reflect the complexity of missingness patterns encountered in real longitudinal data, including intermittent dropout and informative censoring. Positivity violations were approximated via extreme propensity scores, which can also arise from small samples or strong confounding rather than positivity violations per se; exchangeability cannot be directly tested, as it depends on unobserved counterfactuals.[1] The parameter space and the number of levels examined were constrained by computational cost, and each missing data method was implemented with a single, fixed configuration (e.g., MICE with m=5 imputations), which may not represent the full range of implementations used in practice. Finally, complete-case analysis, single mode imputation and hot deck imputation occasionally produced extremely large confidence intervals under small samples, likely reflecting numerical instability in variance estimation, which may have inflated the apparent sample-size effect.

## 5 | Conclusions

Missingness mechanism and choice of missing data method were the dominant determinants of ATE estimation performance under missing data in this simulation study. Multiple imputation by chained equations outperformed complete-case analysis, single mode imputation and stratified hot deck imputation under most conditions, and was the only method to maintain acceptable coverage under MNAR, although all methods showed some deterioration under this mechanism. Positivity violations, approximated via extreme propensity scores, were most pronounced under MNAR, high missing rates and small samples, highlighting the importance of routinely checking propensity score distributions when using propensity-score-based causal methods. Confounding strength primarily affected coverage rather than bias; low-to-moderate confounding was well tolerated by all methods when confounders were fully measured, but strong confounding produced marked reductions in coverage, particularly for complete-case analysis and single mode imputation. Missingness in variables with highly skewed distributions, such as the survival outcomes examined here, degraded the performance of even sophisticated imputation methods, underscoring that variable imbalance can compound the challenges posed by missing data.

These findings support a general preference for multiple imputation over simpler ad hoc methods when handling missing treatment or outcome data in propensity-score-based causal analyses, while highlighting that method performance is highly context-dependent. Future work should extend this comparison to doubly robust and machine-learning-based imputation methods under time-varying confounding, to real-world datasets with more complex or mixed missingness patterns, and to a more systematic exploration of positivity violations and their interaction with missingness mechanisms.

## Conflict of Interest Statement

The authors declare no conflict of interest.

## Data Availability Statement

The simulation code used to generate synthetic data, introduce missingness, apply missing data methods and compute performance measures is available at [].

## Funding

No funding was used in conducting this research.

## Txables

### TABLE 1 Marginal distributions of treatment (A), outcome (Y) and confounder (Z) variables (n=1000, ρ=-0.4)

| Value | A1 | A2 | Y1 | Y2 | Y3 |
|---|---|---|---|---|---|
| 0 | 80.4% | 78.9% | 2.6% | 5.1% | 7.6% |
| 1 | 19.6% | 21.1% | 97.4% | 94.9% | 92.4% |
| Missing | 2.6% | 5.1% | 0% | 0% | 0% |

*Number of comorbidities (Z): 0 = 76.8%; 1 = 10.8%; 2 = 6.7%; 3 = 2.1%; 4 = 3.6%.*

**TABLE 2 Proportion of weights with extreme propensity scores (PS<0.05 or PS>0.95), by parameter and method**

| Parameter | Level | CCA | SMI | HD | MI | Full data |
|---|---|---|---|---|---|---|
| Sample size | 1000 | 0.077 | 0.116 | 0.071 | 0.071 | 0.002 |
| | 5000 | 0.056 | 0.099 | 0.066 | 0.059 | 0.000 |
| Confounding (ρ) | -0.1 | 0.069 | 0.121 | 0.069 | 0.067 | 0.001 |
| | -0.4 | 0.063 | 0.103 | 0.066 | 0.064 | 0.001 |
| | -0.7 | 0.068 | 0.098 | 0.070 | 0.064 | 0.001 |
| Mechanism | MCAR | 0.012 | 0.026 | 0.002 | 0.002 | 0.001 |
| | MAR(A) | 0.003 | 0.093 | 0.002 | 0.004 | 0.001 |
| | MAR(Y) | 0.004 | 0.000 | 0.000 | 0.001 | 0.001 |
| | MNAR | 0.242 | 0.300 | 0.263 | 0.248 | 0.001 |
| Missing rate | 20% | 0.040 | 0.073 | 0.049 | 0.043 | 0.001 |
| | 50% | 0.092 | 0.140 | 0.087 | 0.086 | 0.001 |

*CCA, complete-case analysis; SMI, single mode imputation; HD, stratified hot deck imputation; MI, multiple imputation (MICE).*

**TABLE 3 Partial η² effect sizes from repeated-measures ANOVA of bias, coverage and empirical SE, by parameter and first-order interaction**

| Parameter / interaction | Bias η² | Coverage η² | Empirical SE η² |
|---|---|---|---|
| Mechanism | 0.140* | 0.040 | 0.181* |
| Method | 0.086 | 0.175* | 0.192* |
| Sample size | 0.071 | 0.057 | 0.125 |
| Missing rate | 0.045 | ns | 0.044 |
| Confounding (ρ) | ns | 0.129 | ns |
| Mechanism × method | 0.108 | 0.043 | 0.129 |
| Sample size × mechanism | 0.102 | 0.051 | 0.065 |
| Missing rate × mechanism | 0.102 | ns | 0.069 |
| Sample size × method | 0.046 | 0.022 | 0.063 |
| Missing rate × sample size | 0.029 | ns | 0.011 |
| Missing rate × method | 0.024 | ns | 0.023 |
| ρ × mechanism | ns | 0.082 | ns |
| ρ × sample size | ns | 0.065 | ns |
| ρ × method | ns | 0.045 | ns |
| Missing rate × ρ | ns | 0.014 | ns |

**Denotes practical significance ($\eta^2>0.134$, $p<0.001$). "ns" indicates the effect did not reach statistical significance ($p>0.05$). Full models explained 76% (bias), 74% (coverage) and 90% (empirical SE) of variance.*

**TABLE 4 Tukey HSD comparisons between the extreme levels of each parameter, pooled across methods**

| Parameter | Levels compared | Δ Bias (p) | Δ Coverage (p) | Δ Empirical SE (p) |
|---|---|---|---|---|
| Sample size | n=1000 vs n=5000 | +1.52 (p<0.001) | +0.113 (p<0.001) | +2.84 (p<0.001) |
| Confounding (ρ) | -0.1 vs -0.7 | -0.02 (ns) | -0.200 (p<0.001) | +0.31 (ns) |
| Mechanism | MCAR vs MNAR | -2.67 (p<0.001) | +0.049 (ns) | -4.42 (p<0.001) |
| Missing rate | 20% vs 50% | -1.22 (p<0.001) | +0.006 (ns) | -1.71 (p<0.001) |

*Values are mean differences between the first and second listed level (first - second). ns, not significant (p>0.05).*

**TABLE 5 Best- and worst-performing scenarios for each method and performance measure**

| Method | Measure | Type | Value | Scenario |
|---|---|---|---|---|
| CCA | Bias | best | 0.008 | MCAR, 20%, ρ=-0.1, n=1000 |
| CCA | | worst | 12.77 | MNAR, 50%, ρ=-0.1, n=1000 |
| CCA | Coverage | best | 0.946 | MCAR, 50%, ρ=-0.1, n=5000 |
| CCA | | worst | 0.210 | MCAR, 20%, ρ=-0.7, n=5000 |
| CCA | Empirical SE | best | 0.188 | MCAR, 20%, ρ=-0.4, n=5000 |
| CCA | | worst | 15.32 | MNAR, 50%, ρ=-0.4, n=1000 |
| SMI | Bias | best | 0.020 | MNAR, 20%, ρ=-0.4, n=5000 |
| SMI | | worst | 14.18 | MNAR, 50%, ρ=-0.1, n=1000 |
| SMI | Coverage | best | 0.964 | MCAR, 20%, ρ=-0.1, n=1000 |
| SMI | | worst | 0.044 | MAR(Y), 20%, ρ=-0.7, n=5000 |
| SMI | Empirical SE | best | 0.143 | MAR(Y), 20%, ρ=-0.7, n=5000 |
| SMI | | worst | 16.38 | MNAR, 50%, ρ=-0.4, n=1000 |
| HD | Bias | best | 0.0002 | MAR(A), 20%, ρ=-0.4, n=5000 |
| HD | | worst | 11.54 | MNAR, 50%, ρ=-0.1, n=1000 |
| HD | Coverage | best | 0.962 | MCAR, 20%, ρ=-0.4, n=1000 |
| HD | | worst | 0.042 | MAR(Y), 50%, ρ=-0.7, n=5000 |
| HD | Empirical SE | best | 0.141 | MCAR, 20%, ρ=-0.7, n=5000 |
| HD | | worst | 14.31 | MNAR, 50%, ρ=-0.7, n=1000 |
| MI | Bias | best | 0.005 | MCAR, 50%, ρ=-0.4, n=5000 |
| MI | | worst | 3.05 | MNAR, 50%, ρ=-0.1, n=1000 |
| MI | Coverage | best | 0.984 | MAR(Y), 50%, ρ=-0.1, n=1000 |
| MI | | worst | 0.094 | MAR(Y), 20%, ρ=-0.7, n=5000 |
| MI | Empirical SE | best | 0.161 | MCAR, 20%, ρ=-0.7, n=5000 |
| MI | | worst | 4.83 | MNAR, 50%, ρ=-0.1, n=1000 |

*CCA, complete-case analysis; SMI, single mode imputation; HD, stratified hot deck imputation; MI, multiple imputation (MICE).*

## Figure Legends

FIGURE 1 Directed acyclic graph representing relationships among treatment (A), outcome (Y) and confounder (Z) variables across three treatment time points.

FIGURE 2 Marginal performance measures (absolute bias, 95% CI coverage, empirical SE) by missing data method, for the A1 average treatment effect estimate. CCA, complete-case analysis; SMI, single mode imputation; HD, stratified hot deck imputation; MI, multiple imputation (MICE).

FIGURE 3 Performance measures by sample size (n=1000 vs n=5000) for each missing data method, compared to the full-data model.

FIGURE 4 Performance measures by confounding strength ($\rho$=-0.1, -0.4, -0.7) for each missing data method.

FIGURE 5 Performance measures by missingness mechanism (MCAR, MAR(A), MAR(Y), MNAR) for each missing data method.

FIGURE 6 Performance measures by missing rate (20% vs 50%) for each missing data method.

*Note: Figures are reproduced here for reviewer convenience at reduced resolution. High-resolution figure files (≥600 dpi, .tiff/.eps) will be uploaded separately at submission, per journal requirements.*

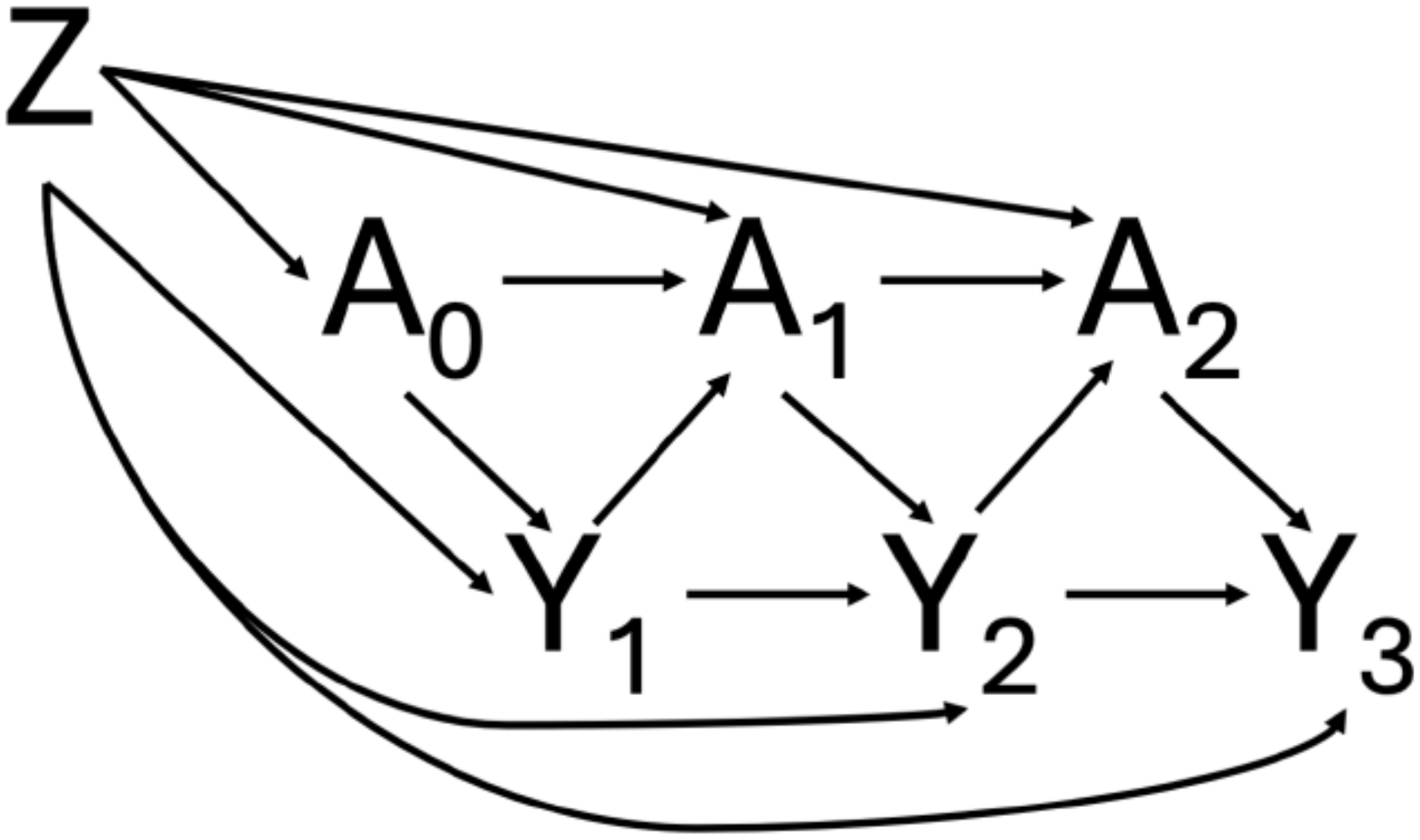

*FIGURE 1 Directed acyclic graph of treatment, outcome and confounder relationships.*

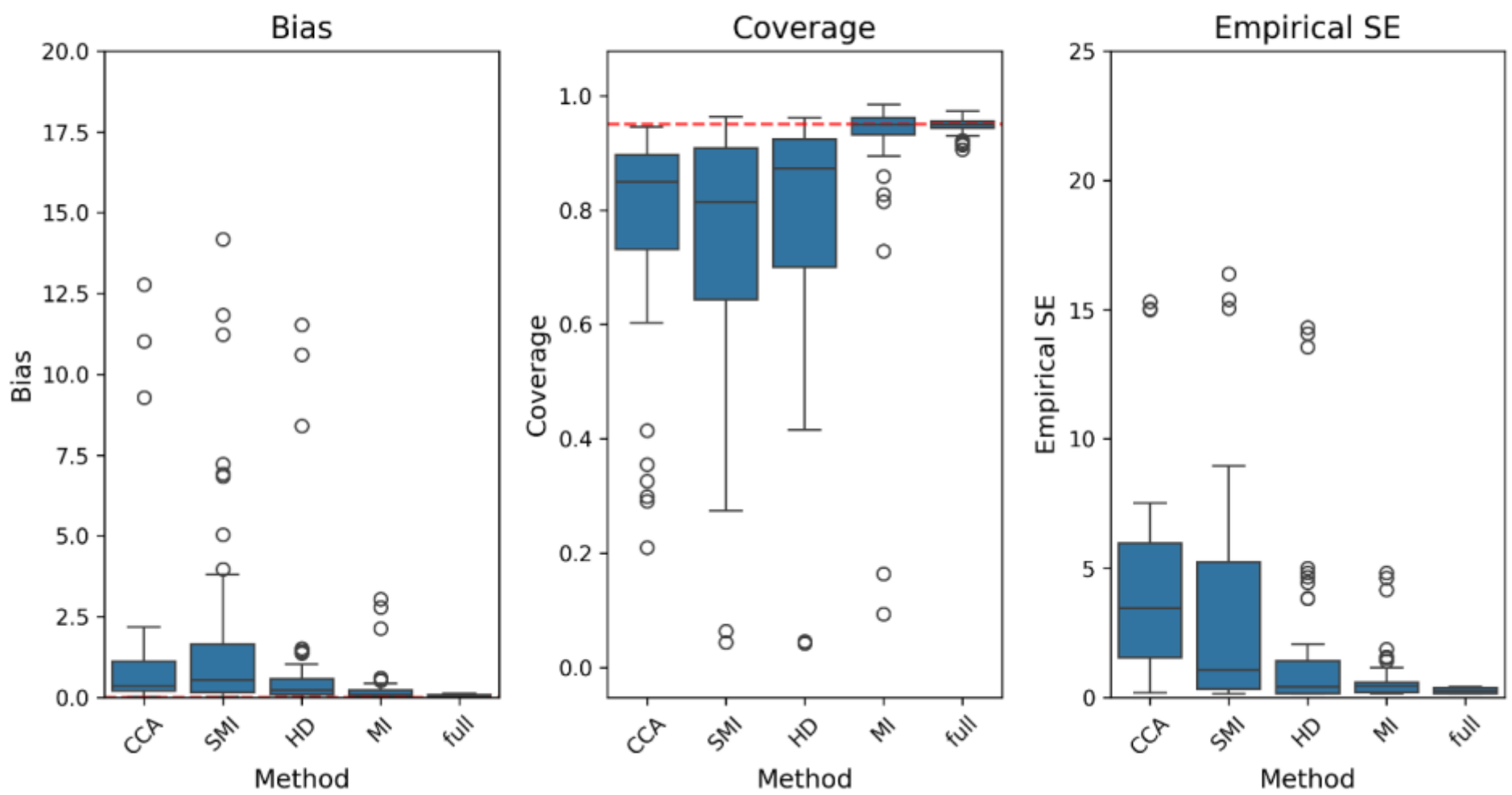

*FIGURE 2 Marginal performance measures by method (ATE estimate A1).*

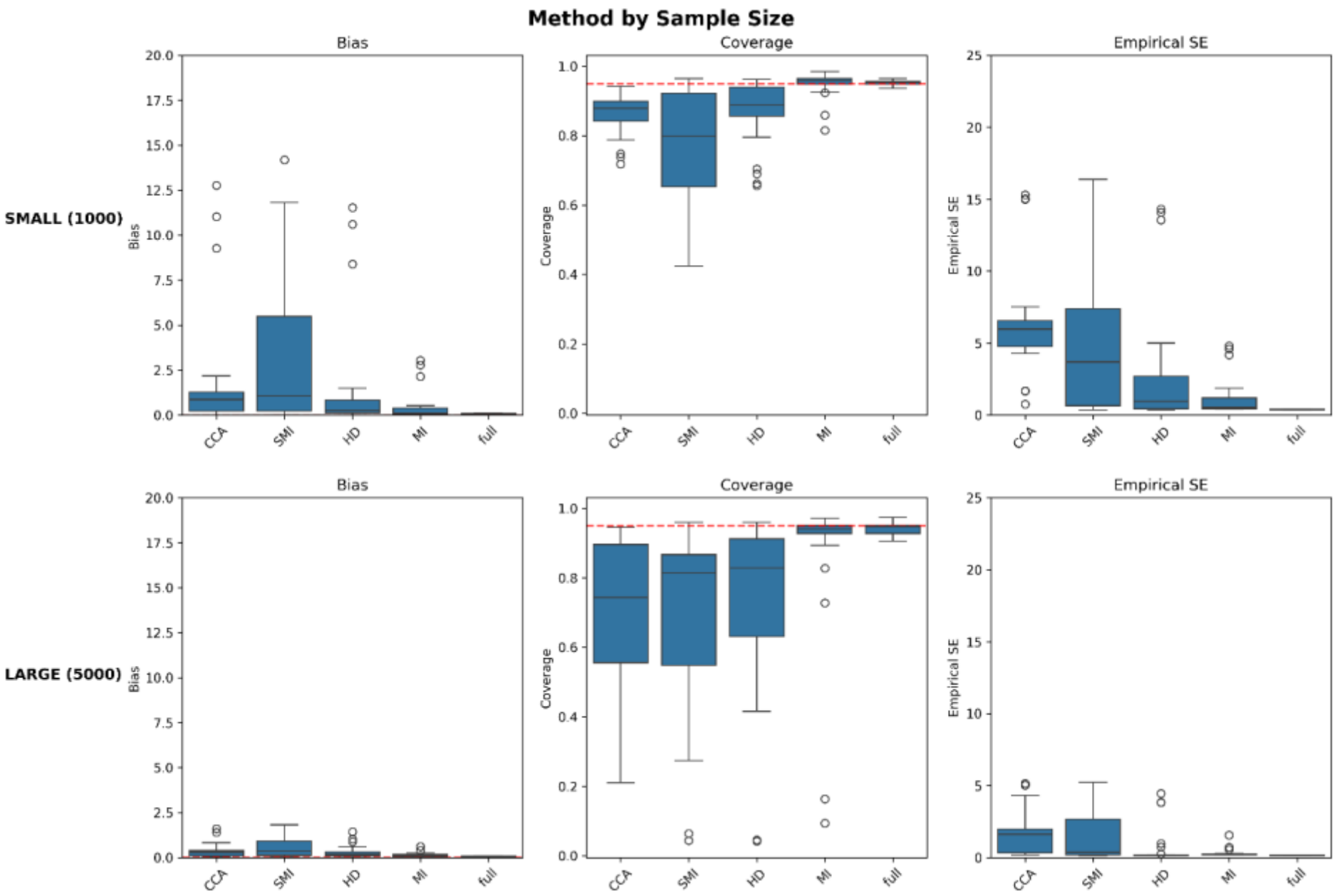

*FIGURE 3 Performance measures by sample size.*

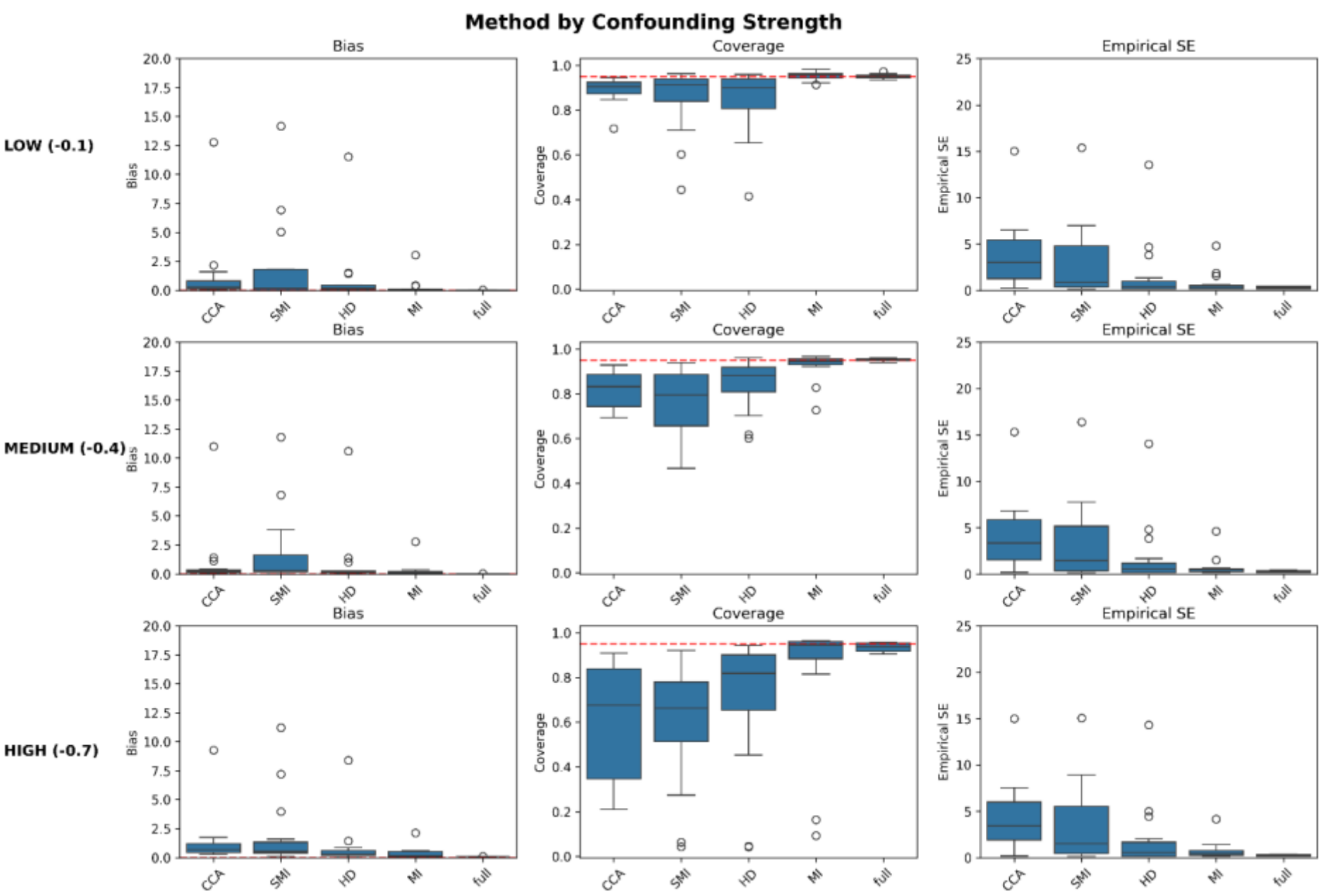

*FIGURE 4 Performance measures by confounding strength.*

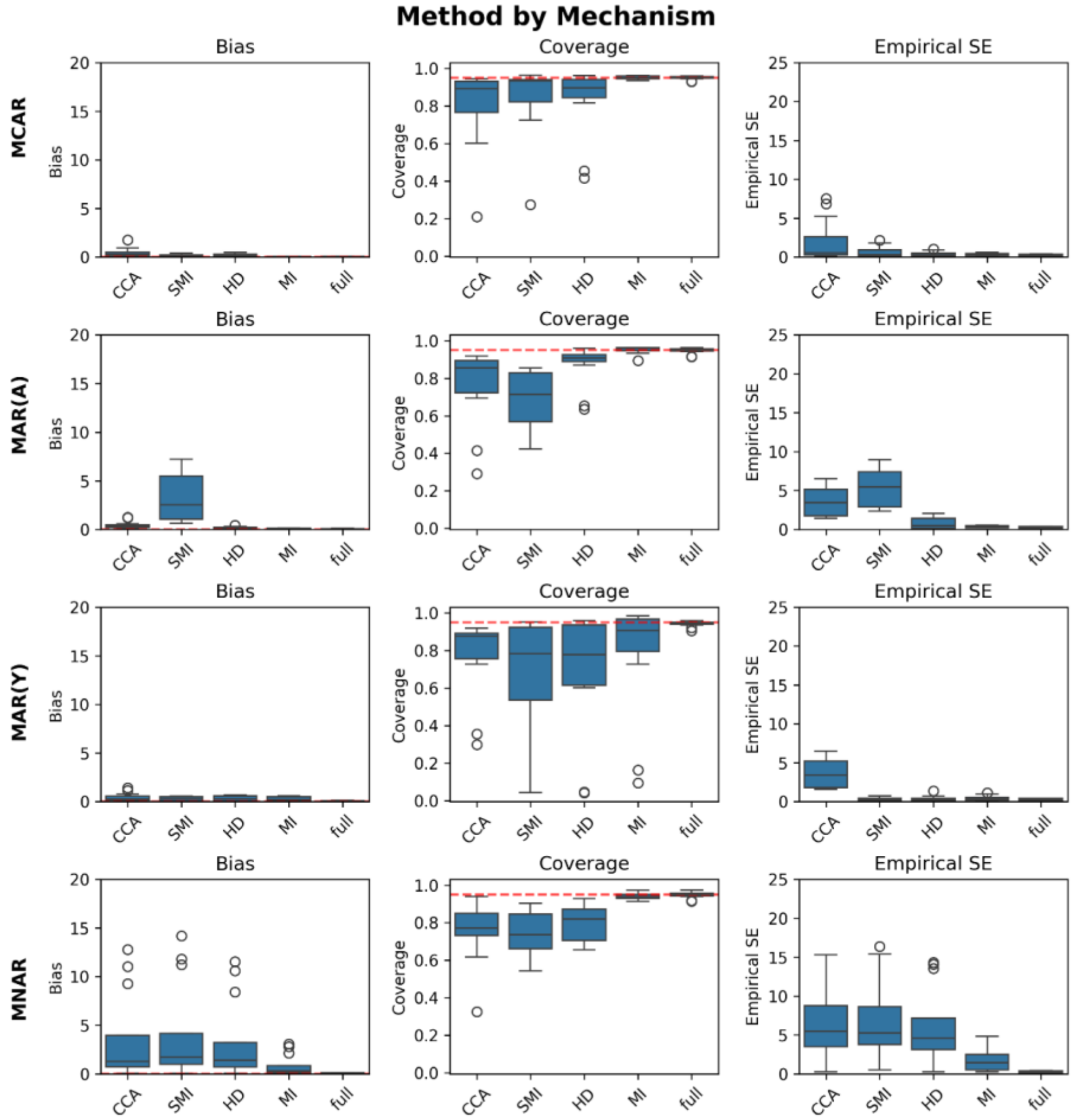

*FIGURE 5 Performance measures by missingness mechanism.*

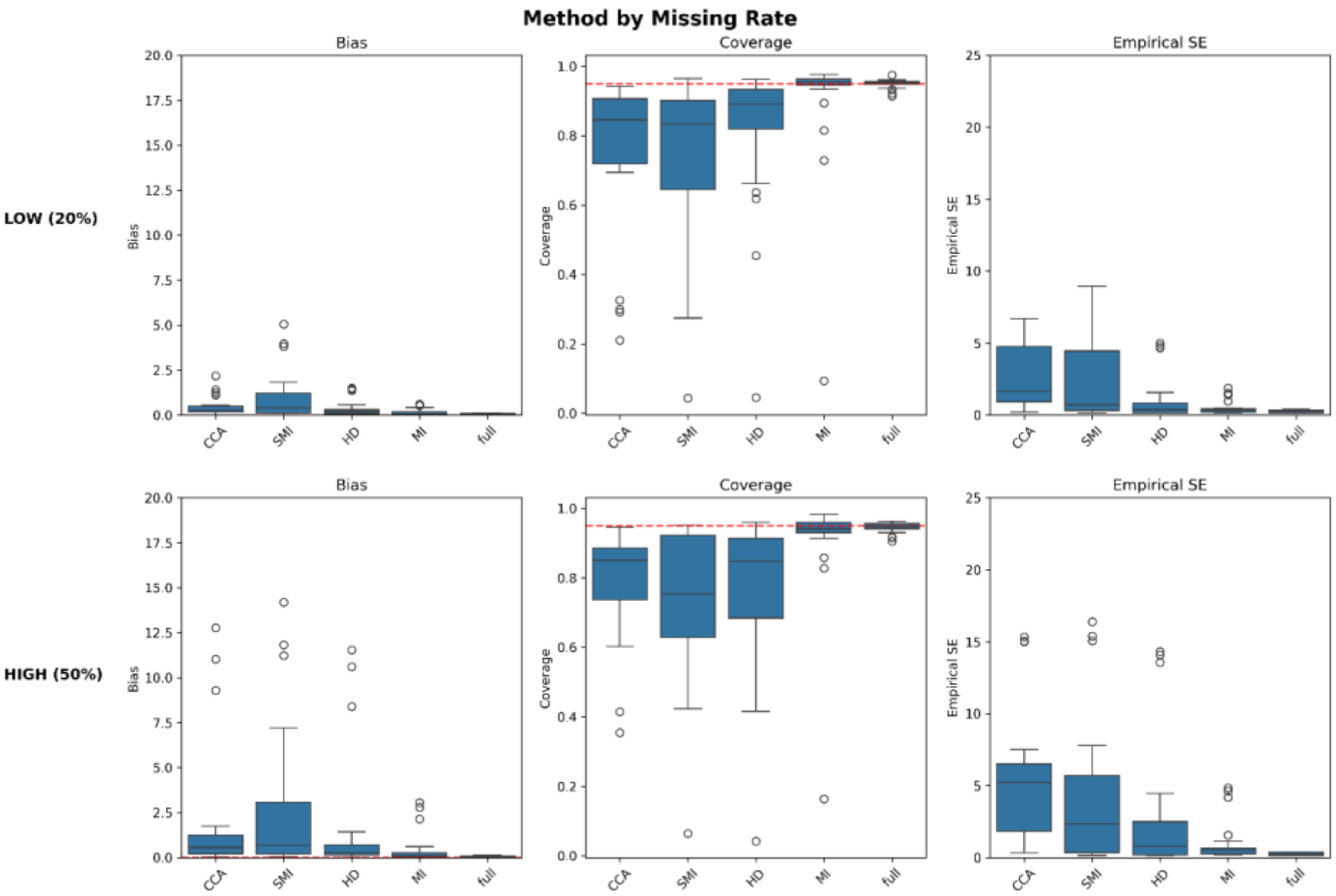

*FIGURE 6 Performance measures by missing rate.*